\documentclass[11pt]{article}

\usepackage{graphics}
\usepackage{amssymb,latexsym}
\usepackage[utf8]{inputenc}
\usepackage{amsmath,amssymb,mathrsfs}
\usepackage{graphicx}
\usepackage{a4wide}
\usepackage[usenames]{color}
\usepackage{cite}%[sort]
\usepackage{caption}
\usepackage{epstopdf}

\usepackage{hyperref}

\newcommand{\non}{\nonumber}
\newcommand{\dszero}{D_{s0}^*}
\newcommand{\dsone}{D_{s1}}
\newcommand{\bszero}{B_{s0}}
\newcommand{\bsone}{B_{s1}}
\newcommand{\lag}{\mathscr L}
\newcommand{\gev}{~\mathrm{GeV}}
\newcommand{\mev}{~\mathrm{MeV}}
\newcommand{\kev}{~\mathrm{keV}}

\newcommand{\al}{&\!\!}
\newcommand{\order}[1]{\mathcal{O}\left(#1\right)}

\begin{document} 

\title{Strong and radiative decays of the $\dszero(2317)$ and $\dsone(2460)$}

\author{Martin~Cleven$^1$\thanks{{\it Email address:} m.cleven@fz-juelich.de}
  , Harald W. Grie{\ss}hammer$^{1,2}$\thanks{{\it E-mail address:}
    hgrie@gwu.edu; permanent address: $2$} , Feng-Kun Guo$^3$\thanks{{\it
      E-mail address:} fkguo@hiskp.uni-bonn.de} ,\\
  Christoph~Hanhart$^{1,4}$\thanks{{\it E-mail address:}
    c.hanhart@fz-juelich.de} \, and Ulf-G.  Mei{\ss}ner$^{1,3,4}$\thanks{{\it
      E-mail address:} meissner@itkp.uni-bonn.de}
  \\[3mm]
  {\small $^1$\it Institut f\"{u}r Kernphysik and J\"ulich Center for Hadron
    Physics},\\
  {\small \it Forschungszentrum J\"{u}lich, D-52425 J\"{u}lich, Germany}\\
  {\small $^2$\it Institute for Nuclear Studies, Department of Physics,} \\
  \small{\it The George Washington University, Washington DC 20052, USA}\\
  {\small $^3$\it Helmholtz-Institut f\"ur Strahlen- und Kernphysik and
    Bethe Center for Theoretical Physics,}\\
  {\small \it  Universit\"at Bonn, D-53115 Bonn, Germany}\\
  {\small$^4$\it Institute for Advanced Simulation,
    Forschungszentrum J\"{u}lich, D-52425 J\"{u}lich, Germany}\\
}

\maketitle
\begin{abstract}
  Since their discovery in 2003, the open charm states $\dszero(2317)$ and
  $\dsone(2460)$ provide a challenge to the conventional quark model. In
  recent years, theoretical evidence has been accumulated for both states in
  favor of a predominantly $DK$ and $D^*K$ molecular nature, respectively.
  However, a direct experimental proof of this hypothesis still needs to be
  found. Since radiative decays are generally believed to be sensitive to the
  inner structure of the decaying particles, we study in this work the
  radiative and strong decays of both the $\dszero(2317)$ and $\dsone(2460)$,
  as well as of their counterparts in the bottom sector.  While the strong
  decays are indeed strongly enhanced for molecular states, the radiative
  decays are of similar order of magnitude in different pictures. Thus, the
  experimental observable that allows one to conclusively quantify the
  molecular components of the $\dszero(2317)$ and $\dsone(2460)$ is the
  hadronic width, and not the radiative one, in contradistinction to
    common belief. We also find that radiative decays of the sibling states
  in the bottom sector are significantly more frequent than the hadronic ones.
  Based on this, we identify their most promising discovery channels.
\end{abstract}

\thispagestyle{empty}

\newpage
 
\section{Introduction}
Since the beginning of this millenium, mounting experimental evidence in
hadronic spectroscopy puts into question quark models like the Godfrey-Isgur
model~\cite{Godfrey:1985xj} that successfully described the ground and some
low excited states of mesons with open charm or bottom. This picture was
challenged when two narrow resonances with open charm were discovered by the
BaBar~\cite{Aubert:2003fg} and CLEO collaborations~\cite{Besson:2003cp},
respectively. These states are now named $\dszero(2317)$ and $\dsone(2460)$
and referred to in the following as $\dszero$ and $\dsone$, respectively.
Their respective masses were about $160\mev$ and $70\mev$ below the
predictions of the Godfrey-Isgur quark model.  On the other hand, the states
are located by almost the same amount of about 45~MeV below the $DK$ and
$D^*K$ thresholds, respectively. This appears a mere numerical coincidence
in quark models, and is a consequence of the parity doubling assumption in
Refs.~\cite{Nowak:1992um,Bardeen:2003kt,Nowak:2003ra}.  However, as stressed
in Ref.~\cite{Guo:2009id}, this can be explained naturally if the systems are
bound states of the $DK$ and $D^*K$ meson pairs,
respectively~\cite{Barnes:2003dj,vanBeveren:2003kd,vanBeveren:2003jv,
  Kolomeitsev:2003ac, Guo:2006fu,Zhang:2006ix,Guo:2006rp}.

Weinberg introduced a model-independent way to quantify the molecular
admixture in the wave function of a physical state~\cite{Weinberg:1962hj}.
The relation between the coupling constant $g$ of a hadronic molecule with a
mass $M$ and a binding energy $\epsilon=m_1+m_2-M$ to its constituents with
masses $m_1$ and $m_2$ and the reduced mass $\mu=m_1m_2/(m_1+m_2)$ is found to
be
\begin{equation}\label{eq:weinberg}
 g^2=16\pi\lambda^2\frac{(m_1+m_2)^2}{\mu}\sqrt{2\mu \epsilon}  + 
\order{R\sqrt{2\mu \epsilon}}\ ,
\end{equation}
where $1/R$ denotes the momentum scale related to dynamics not included
explicitly, such as effective range corrections or other channels.  The
parameter $\lambda^2$ is the probability of finding a two-body continuum state
in the physical state. It is thus zero for an elementary particle and one for
a pure two-body molecule. For a shallow bound state whose binding energy is
small so that $R\sqrt{2\mu\epsilon}\ll1$, the pole contribution dominates the
$S$-matrix elements in the near-threshold region, in particular the scattering
length. This makes $g$ in principle accessible to experiment, albeit $DK$
scattering is not likely to be directly observed experimentally in the near
future. However, the scattering properties can be calculated using lattice
quantum chromodynamics (QCD).  Indeed, there have been lattice calculations of
the $S$-wave isoscalar $DK$ scattering length. It was calculated directly in
Ref.~\cite{Mohler:2013rwa,Lang:2014yfa} at two pion masses, and the obtained
values agree with the ones from the indirect calculation in
Ref.~\cite{Liu:2012zya}.% ~\footnote{
In lattice QCD, the isoscalar $DK$ scattering is relatively difficult because
of the presence of the disconnected Wick contractions which are of leading
order at both the $1/N_c$ expansion and chiral expansion~\cite{Guo:2013nja}.
In Ref.~\cite{Liu:2012zya}, the charmed meson-light meson scattering lengths
for the channels which are free of disconnected contractions are calculated,
and then the $DK$ scattering length was extracted~\cite{Guo:2009ct} using
unitarized chiral perturbation theory, the parameters of which were determined
from fitting to the lattice results. In this sense, we refer to the
calculation of the $DK$ scattering length as ``indirect''. These lattice results agree
perfectly with the prediction of Eq.~(\ref{eq:weinberg}) for $\lambda=1$
taking into account the uncertainties.  This provides strong evidence from the
theoretical point of view that the $\dszero$ and $\dsone$ are $D^{(*)}K$
molecular states. However, a clear experimental proof is still missing.

It was stressed in 
Refs.~\cite{Hofmann:2003je,Faessler:2007gv,Guo:2008gp, Liu:2012zya} that the 
leading loop contributions to the hadronic widths of  $\dszero$ and $\dsone$ are quite
sensitive to $g^2$ and thus allow one to quantify their molecular admixtures
experimentally. Especially,  no additional 
counterterm is present at leading order~(LO).  
The 
situation for the radiative decays is less clear. While
Refs.~\cite{Gamermann:2007bm,Faessler:2007gv,Faessler:2007us} provide
predictions, a LO counterterm obstructs a prediction in
Ref.~\cite{Lutz:2007sk} and hampers the sensitivity to the coupling constant 
$g^2$.

In this work, we reinvestigate the decays of the $\dszero$ and $\dsone$ in an
effective field theory description appropriate for these systems. Our key
finding is that the radiative decays of the $\dszero$ and $\dsone$ are
insensitive to their precise nature, contrary to  the strong decays. In 
particular,
the electromagnetic transition rates are not enhanced by the fact that for
molecular states the decay has to run via meson loops because of the presence
of counterterms.  Using a different formalism, we also confirm the findings of
Ref.~\cite{Lutz:2007sk} that radiative decays suffer from a contact interaction
with unknown coefficient already at LO.  This supports the claim
that only an experimental hadronic width of the $\dszero$ and $\dsone$ of the
order of $100$~keV can be regarded as the smoking gun for a predominantly
molecular nature of these two states.

Unfortunately, the experimental information available at present is rather
limited. At best, ratios between hadronic and radiative decays are published,
with only upper limits for most transitions. There is hope that with the
advent of high precision and high intensity experiments like 
$\overline{\text{P}}$ANDA, this situation will be improved significantly.

Since heavy quark flavor symmetry connects the open charm and bottom sectors,
states similar to the $\dszero$ and $\dsone$ are expected in the open bottom
sector.  Such predictions have been made for conventional mesons with parity
doubling~\cite{Bardeen:2003kt}, using heavy quark effective 
theory~\cite{Mehen:2005hc,Cheng:2014bca} or for $B^{(*)}K$ molecules in a 
variety of publications~\cite{Kolomeitsev:2003ac,Guo:2006fu,Zhang:2006ix,
Guo:2006rp,Cleven:2010aw}.
Here, we update the latter class of works, and especially identify radiative
decays as the probably most promising discovery modes of the bottom-partners
of the $\dszero$ and $\dsone$.

The paper is organized as follows. The theoretical framework and the 
interaction Lagrangians are presented in Sec.~\ref{sec:framework}. Both the 
isospin breaking hadronic decays and radiative decays of the $\dszero$ and 
$\dsone$ and their bottom partners are calculated in 
Sec.~\ref{sec:twobodydecays}. The last section contains a brief summary.

%%%%%%%%%%%%%%%%%%%%%%%%%%%%%%%%%%%%%%%%%%%%%%%%%%%%%%%%
%
\section{Framework}\label{sec:framework}
%
%%%%%%%%%%%%%%%%%%%%%%%%%%%%%%%%%%%%%%%%%%%%%%%%%%%%%%%%
Various earlier works demonstrated that both the $\dszero$ and $\dsone$ can
straightforwardly be produced by unitarizing $DK^{(*)}$ scattering amplitudes
which are derived, for example, from chiral perturbation theory (CHPT) at LO
~\cite{Kolomeitsev:2003ac,Guo:2006fu,Guo:2006rp,Gamermann:2006nm,
  Gamermann:2007fi} or next-to-leading order
(NLO)~\cite{Hofmann:2003je,Lutz:2007sk,Guo:2008gp,Guo:2009ct,Liu:2012zya}.  
These amplitudes were also used to calculate their strong decays. In principle, 
the electromagnetic decays could also be addressed with the full set of 
equations
by gauging the integral equation~\cite{Mai:2012wy}. However, since we are
interested in an observable close to the resonance pole only, we can take a
simpler route. First, we extract the pole residues from the
full calculation, and then use these as input of a one-loop evaluation of the
actual decays. 
For a proper field theoretical derivation of the
  connection between the two approaches in a different context, see Sec.~3.3
  of Ref.~\cite{Meissner:2005bz}.

Our approach is based on the Lagrangian describing the coupling of the
molecules to a heavy-light meson pair in an $S$-wave: 
\begin{eqnarray}\nonumber
\lag_\mathrm{Mol}^{D} \al=\al 
g_{DK}D_{s0}^*\left(D^{+\dagger}K^{0\dagger}+D^{0\dagger}K^{+\dagger} 
\right)+g_{D_s\eta}D_{s0}^*D_s^{\dagger}\eta^\dagger\\
 \al\al +g_{D^*K}D^\mu_{s1}\left(D^{*+\dagger}_\mu 
K^{0\dagger}+D^{*0\dagger}_\mu 
K^{+\dagger}\right)+g_{D^*_s\eta}D^\mu_{s1}D_{s,\mu}^{*\dagger}\eta^\dagger
+\mathrm{h.c.},\label{eq:LagrMol}
\end{eqnarray}
where $g_i$ denote the corresponding coupling constants. Since we are only 
interested in the near-threshold region, a constant coupling is used for the 
$S$-wave coupling.
In Ref.~\cite{Liu:2012zya}, the $\dszero$ pole was generated dynamically using
unitarized NLO heavy meson-Goldstone boson scattering amplitudes. The
low-energy constants (LECs) were fit to lattice calculations for various 
scattering
lengths. The same values of the LECs are used in this work.
We here mainly present the extension of the earlier formalism necessary for this work.
For more details, we refer to 
Refs.~\cite{Guo:2009ct,Liu:2012zya,thesis}. Each of the isoscalar heavy 
meson-kaon scattering amplitudes has a pole below threshold which 
corresponds to the particle of interest. The coupling constants defined in the 
Lagrangian in Eq.~\eqref{eq:LagrMol} are then
determined from the residues of these poles:
\begin{eqnarray}\label{eq:couplingsCharm}
 &&g_{DK}=(9.0\pm0.5)\gev,\qquad g_{D^*K}=(10.0\pm0.3)\gev \non \\
 &&g_{D_s\eta}=(8.0\pm0.2)\gev,\qquad g_{D^*_s\eta}=(7.7\pm0.5)\gev,
\end{eqnarray}
where the uncertainties are propagated from the errors of the LECs with 
correlations taken into account.
The couplings of the $\dszero$ ($\dsone$) to the $DK$ ($D^*K$) turn out to be
larger than those to the $D_s\eta$ ($D_s^*\eta$). In addition, the $DK$ channel
indeed dominates over the $D_s\eta$ channel, thanks to an enhancement by a
factor of
\begin{equation}
  {\sqrt{(m_\eta+m_{D_s}-M)/(m_D+m_K-M)}\simeq 2}\;.
\end{equation}
Accordingly, applying Eq.~(\ref{eq:weinberg}) to the couplings in
Eq.~(\ref{eq:couplingsCharm}), we find values of $\lambda^2$ for both  $\dszero$ and 
$\dsone$ 
of about $0.8$, however, with a sizable uncertainty of the order
of 50\%, which comes from $R\sim1/\sqrt{2\mu(m_\eta+m_{D_s}-M)}$, due to the proximity of the $D_s\eta$ 
channel---this provides additional evidence
for the interpretation of  $\dszero$ and $\dsone$ as predominantly
molecular states.
Note that the mentioned large uncertainty refers to
quantifying the molecular component of the scalar and axial-vector states; the residues themselves are known to
much higher accuracy, see Eq.~\eqref{eq:couplingsCharm}, and it is their uncertainty that matters
for the calculations below. 

\begin{table}[t]
  \centering \caption{Comparison of our predictions of the masses of the
    ${\bar B}K$ and ${\bar B}^*K$ bound states with those in 
    Refs.~\cite{Kolomeitsev:2003ac,Guo:2006fu,Guo:2006rp,Mehen:2005hc,
      Bardeen:2003kt}. All masses are given in MeV.
      \medskip } 
  \renewcommand{\arraystretch}{1.3}
\begin{tabular}{|l|c | c | c | c | c | c|} 
\hline\hline
		& This calculation	 & Ref.~\cite{Kolomeitsev:2003ac}	& 
Refs.~\cite{Guo:2006fu,Guo:2006rp} & Ref.~\cite{Mehen:2005hc}  & 
Ref.~\cite{Cleven:2010aw}	& Ref.~\cite{Bardeen:2003kt}	\\ \hline\hline
 $M_{B^*_{s0}}$& $5625\pm45$		& $5643$ 				& $5725\pm39$ 				& $5667$ & $5696 \pm 40$ 		& $5718 \pm 35$		\\ \hline
 $M_{B_{s1}}$  & $5671\pm45$		& $5690$ 				& $5778\pm7$  				& $5714$	& $5742 \pm 40$	& $5765 \pm 35$		\\ \hline\hline
\end{tabular}
\label{tab:BMasses}
\end{table}

Since heavy quark flavor symmetry allows us to use the same parameters and
predict the heavy-flavor partners, we can extend these calculations to the
open bottom sector. In our previous study~\cite{Cleven:2010aw}, we took the
same subtraction constant which is used to regularize the divergent two-meson 
loop integrals in dimensional regularization for both the bottom and charm 
systems. Now, we use a
different method which makes the transmission of the scale-dependence of the 
loop integrals more transparent/physical: we first use a three-momentum sharp 
cutoff to regularize the loop integral and fix it to reproduce the 
dimensional-regularized loop in the charm sector, and use the same cutoff 
to determine the value of the subtraction constant in the bottom sector. Then 
the masses of the generated states with positive parity can be calculated by 
searching for poles of the scattering amplitudes. The results are presented in 
the first column of
Table~\ref{tab:BMasses}.  The uncertainty contains both that of the LECs 
and of the heavy-flavor symmetry breaking, added in quadrature. We
estimated the latter as $ (\Lambda_\text{QCD}/m_c)\epsilon\sim 40\mev$.  Within
uncertainties, the masses obey the relation
\begin{equation}
 M_{\bsone}-M_{\bszero}\simeq M_{B^*}-M_{B}.
\end{equation}
In Table~\ref{tab:BMasses}, we also compare our results with previous studies. Our
values agree within errors with
Refs.~\cite{Kolomeitsev:2003ac,Mehen:2005hc,Cleven:2010aw}, while there is
some discrepancy to the results of
Refs.~\cite{Guo:2006fu,Guo:2006rp,Bardeen:2003kt}.

The Lagrangian for coupling the bottom molecules to heavy-light meson pairs 
is analogous to Eq.~(\ref{eq:LagrMol}).  The corresponding residues for the $B_{s0}^*$ and $B_{s1}$
read
\begin{eqnarray}\label{eq:couplingsBottom}
  &&g_{BK}=(30\pm 1)\gev,\qquad g_{B^*K}=(30\pm 1)\gev \\
  &&g_{B_s\eta}=(12\pm 6)\gev,\qquad g_{B^*_s\eta}=(10\pm 7)\gev.
\end{eqnarray}
The larger couplings reflect the fact that the bottom states are more deeply
bound, as expected from Eq.~(\ref{eq:weinberg}). Indeed, the large binding
energy renders useless any estimate of the probability as $\lambda^2$ via
Eq.~(\ref{eq:weinberg}). 

To calculate the radiative decays, we need the magnetic moments of the heavy
mesons in addition to the electric photon-meson coupling which comes from
gauging the kinetic term of the heavy mesons.  The Lagrangian
reads~\cite{Amundson:1992yp,Hu:2005gf}~\footnote{Notice that in our notation,
  the presence of the factor $M_H$ renders the fields $P_a$ and $P_{a\mu}^*$
  to have an energy dimension 1. This is different from
  Refs.~\cite{Amundson:1992yp,Hu:2005gf} where the dimension of these fields
  is $3/2$.}
\begin{eqnarray}\label{LagrMagnMom}
 \lag_{\rm mag.~mom.}&{=}&\frac i2 e
 F_{\mu\nu}M_H\left[\varepsilon^{\mu\nu\alpha\beta}v_\alpha\left(\beta
     Q+\frac{Q'}{m_Q}\right)_{ab}\left(P_aP^{*\dagger}_{b\beta}-P^{*}_{a\beta}P^\dagger_b \right) \right.\\ 
 &&\left.+P_a^{*\mu} P^{*\dagger\nu}_b\left(\beta Q-\frac{Q'}{m_Q}\right)_{ab}
 \right]\non 
\end{eqnarray}
where $M_H$ is the mass of the heavy meson, and the pseudoscalar (vector)
mesons with open charm are collected in $P_a$ ($P^*_{a\mu}$) with $a$ labeling 
the light flavors
\begin{eqnarray}
 P=\left(D^0,D^+,D_s^+\right),\quad
 P^*_{\mu}=\left(D^{*0}_{\mu},D^{*+}_{\mu},D_{s,\mu}^{*+}\right). 
\end{eqnarray}
The $\beta Q$ term, where
${Q=\mathrm{diag}(2/3,-1/3,-1/3)}$ is the light quark charge matrix, comes 
from the magnetic moment of the light degrees of freedom, and
the $Q'/m_Q$ term is the magnetic moment coupling of the heavy quark with $Q'$
and $m_Q$ being the charge and mass of the heavy quark, respectively. The 
quantities $\beta$ and $m_Q$ can be
fixed from experimental data for $\Gamma(D^{*0}\to D^0\gamma)$ and
$\Gamma(D^{*+}\to D^+\gamma)$. We will use one set of values 
determined in Ref.~\cite{Hu:2005gf}, which are
$ 1/\beta=379\mev$ and $ m_c=1863\mev.$
The transition to the bottom sector is made by using $m_b=4650\mev$ and the 
same value for $\beta$.

\begin{figure}[t]
\centering
\includegraphics[width=.8\linewidth]{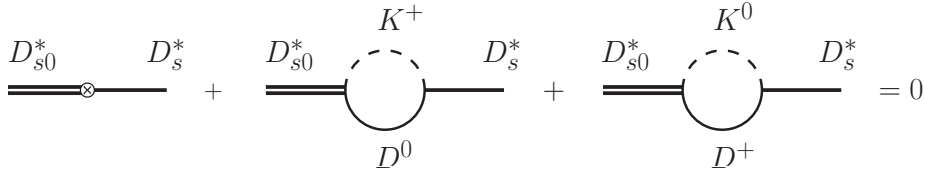}
\caption{\label{fig:mixing} The mass renormalization mechanism that ensures
  that the $\dszero$ and $D_s^*$ do not mix for the physical particles. The
  loops are evaluated at $p^2=M_{\dszero}^2$.  Double lines denote 
  molecules, single lines charmed mesons, and dashed lines kaons.}
\end{figure}

Since the longitudinal components of the vector fields, $\partial_\mu
P_a^{*\mu}$, have scalar quantum numbers, hadronic loops couple them to the
scalar fields.  In this way, the longitudinal components of the vector fields
contribute to the self-energy of the scalar field. Analogously, the
longitudinal components of the axial vectors couple to the pseudoscalar
fields.  Thus, for the purpose of renormalization, we have to add the
counterterms
\begin{eqnarray}\label{eq:LagrMR}
 \mathscr L_\text{long.}=C_{DK} D_{s0}^*\left(\mathcal D_\mu 
D_s^{*\dagger\mu}\right)-C_{D^*K}\left[\left(\mathcal D_\mu 
D_{s1}^{*\mu}\right)D_s^\dagger-i\varepsilon_{\mu\nu\alpha\beta}v^\alpha 
D_{s1}^\mu D_s^{*\dagger\beta} \right] \ ,
\end{eqnarray}
with the coupling constants $C_{DK}$ and $C_{D^*K}$ adjusted to cancel the
loops at the poles, cf.~Fig.~\ref{fig:mixing} and
Ref.~\cite{Gamermann:2007bm}. 

Furthermore, the Lagrangian for the leading contact interactions for the 
radiative decays reads
\begin{eqnarray}
 \lag_{\rm Contact} \al=\al\kappa F_{\mu\nu}\left(v^\mu \dszero 
D_s^{*\dagger\nu}+\dsone^\mu v^\nu 
D_s^\dagger+\varepsilon^{\mu\nu\alpha\beta}{\dsone}_\alpha 
D^{*\dagger}_{s\beta}\right) \nonumber\\
 \al\al 
 + \tilde\kappa 
\varepsilon^{\mu\nu\alpha\beta}F_{\mu\nu}v_\beta D_{s1,\alpha} 
D_{s0}^{*\dagger} + \text{h.c.} \label{eq:contact1}
\end{eqnarray}
We will discuss the relative importance of contact interactions and loop
diagrams in a CHPT power counting scheme in
Sec.~\ref{sec:radiative}.

For numerical calculations, we will take the following values for the meson 
masses~\cite{Beringer:1900zz}:
\begin{equation}\label{eq:masses} 
\begin{array}{lll}
 M_{D^0}=1864.86 \mev,&   M_{D^+ }=1869.62 \mev,& M_{D_s^+ }= 1968.49\mev, 
\non\\
 M_{D^{*0}}= 2006.98\mev,& M_{D^{*+}}=2010.28 \mev,&  M_{D_s^{*+} }= 
2112.3\mev, \non\\
 M_{B^+}=5279.25 \mev,&   M_{B^0 }= 5279.58\mev,& M_{B^0 }= 5366.77 \mev, \non\\
 M_{B^{*+} }=5325.2\mev,&  M_{B^{*0} }= 5325.2 \mev,&   M_{B_s^{*0} }= 
5415.4\mev, \non\\
 M_{\pi^0} = 134.98~{\rm MeV},&  M_{\pi^+}=139.570\mev,& M_{\eta} = 547.85~{\rm 
MeV}, \non\\
 M_{K^+}= 493.677\mev,& M_{K^0}=497.614\mev.&
\end{array}
\end{equation}
It is important to also specify the uncertainties of the mass differences used
in our approach: $M_{D^+}-M_{D^{0}}=(4.8\pm0.2)\mev$, $
M_{D^{*+}}-M_{D^{*0}}=(3.3\pm0.2)\mev$, $
M_{B^+}-M_{B^{0}}=(-0.33\pm0.24)\mev$, $
M_{B^{*+}}-M_{B^{*0}}=(-0.33\pm0.24)\mev$~\cite{Beringer:1900zz}. The mass
splittings in the charm and bottom sectors have different patterns because the
interference between the $m_d-m_u$ contribution and the electromagnetic
contribution is different~\cite{Guo:2008ns}.

%%%%%%%%%%%%%%%%%%%%%%%%%%%%%%%%%%%%%%%%%%%%%%%%%%%%
%
\section{Two-Body Decays}\label{sec:twobodydecays}
\subsection{Hadronic Decays}\label{sec:hadronic}
%
%%%%%%%%%%%%%%%%%%%%%%%%%%%%%%%%%%%%%%%%%%%%%%%%%%%%

In this section, we calculate the hadronic decay widths $\dszero\to D_s\pi^0$ 
and $\dsone\to D_s^*\pi^0$ and their corresponding bottom partners. The narrow widths of the charmed states can only be understood,
 if  they are isoscalar states for then the pionic decays violate isospin. One natural 
decay mechanism, which is present irrespective of the assumed nature of the 
states, is the strong decay of the scalar (axial vector) state into a $D_s$ 
($D_s^*$) and a virtual $\eta$-meson, followed by the isospin violating 
transition to a pion via the $\pi^0$-$\eta$ mixing amplitude
$\epsilon_{\pi\eta}=0.013\pm0.001$. This amplitude is analytic in the quark
masses and scales as $(m_u-m_d)/m_s$. Different groups using different
underlying models for the $D_{sJ}$ states report hadronic widths due to the 
$\pi^0$-$\eta$ mixing ranging between 3 and 25
keV~\cite{Bardeen:2003kt,Guo:2006fu,Guo:2006rp,Lutz:2007sk,Faessler:2007gv,Guo:2008gp}.
 
 For molecular states
one more isospin violating mechanism exists, since they  decay through 
meson loops~\cite{Lutz:2007sk,Faessler:2007gv,Guo:2008gp}.  
Because the charged and neutral mesons have different masses, the meson loops 
of, for instance, $D^+K^0$ and $D^0K^+$ are different numerically. 
This difference introduces an additional, sizable isospin breaking that is specific
for molecular states.
In effect, most studies found hadronic widths for the
hadronic molecules larger than $100$~keV. Especially, the most refined
investigation including lattice data to fix higher-order operators finds a
hadronic width of $(133\pm 22)$ keV~\cite{Liu:2012zya}.

%%%%%%%%%%%%%%%%%%%%%%%%%%%%%%%%%%%%%%%%%%%%%%%%%%%%
\begin{figure}%[htbp]
\centering
%\captionsetup{width=.88\textwidth}
\includegraphics[width=.8\linewidth]{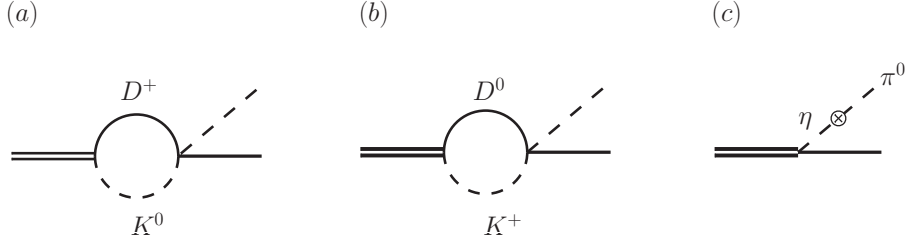}
\caption{\label{fig:hadronic} The two mechanisms that contribute to the
  hadronic width of the $\dszero$. (a) and (b) represent the 
  nonvanishing difference
  for the loops with $D^+K^0$ and $D^0K^+$, respectively. (c) depicts the 
  decay via $\pi^0$-$\eta$ mixing.}
\end{figure}
%%%%%%%%%%%%%%%%%%%%%%%%%%%%%%%%%%%%%%%%%%%%%%%%%%%%
%

%%%%%%%%%%%%%%%%%%%%%%%%%%%%%%%%%%%%%%%%%%%%%%%%%%%%
\begin{table}[t]%[htbp]
\centering
\caption{Hadronic decay widths from different mechanisms.
\medskip}
\renewcommand{\arraystretch}{1.3}
\begin{tabular}{|l|c|c|c|}
\hline\hline
	Decays & loops 	& $\pi^0$-$\eta$ mixing	& full result \\ 
\hline\hline
$\dszero\to D_s\pi^0$ 	& $(26 \pm 3)\kev$		& $(23 \pm 3)\kev$	& $(96 \pm 19)\kev$ \\ \hline
$\dsone\to D_s^*\pi^0$	& $(20 \pm 3)\kev$		& $(19 \pm 3)\kev$	& $(78\pm14)\kev$ \\ \hline
$\bszero\to B_s\pi^0$ 	& $(8 \pm 5)\kev$		& $(6 \pm 6)\kev$	& $(0.8 \pm 0.8)\kev$ \\ \hline
$\bsone\to B_s^*\pi^0$ 	& $(7 \pm 4)\kev$		& $(5 \pm 5)\kev$	& $(1.8 \pm 1.8)\kev$ \\ \hline\hline
\end{tabular}
\label{tab:hadronic}
\end{table}

%%%%%%%%%%%%%%%%%%%%%%%%%%%%%%%%%%%%%%%%%%%%%%%%%%%%
%
% Results
%
%%%%%%%%%%%%%%%%%%%%%%%%%%%%%%%%%%%%%%%%%%%%%%%%%%%%

In our formalism, the isospin violating decays are represented by the diagrams
shown in Fig.~\ref{fig:hadronic}. For the calculation of the loop diagrams, we
put the heavy-light rescattering vertex on shell for two reasons. First, the
off-shell effects contribute to subleading orders only. Second, we followed
the same procedure when solving the scattering equations that led to the
values of the residues in Eqs.~(\ref{eq:couplingsCharm}) and
(\ref{eq:couplingsBottom}). the results are  given in
Its first column lists the
results from the meson loops, see Fig.~\ref{fig:hadronic} (a) and (b). Since
the mass differences in the $D$ sector are significantly larger than in the
see above,  the contribution of the loops to the widths is
significantly larger for the former than for the latter. The uncertainties are
propagated from the uncertainties of the residues in
Eqs.~(\ref{eq:couplingsCharm}) and (\ref{eq:couplingsBottom}), as well as
those of the LECs for the heavy-light rescattering vertex. When the widths
are calculated exclusively from $\pi^0$-$\eta$-mixing, diagram (c) of
Fig.~\ref{fig:hadronic}, we find around $20\kev$ in the charm sector and about
$5\kev$ in the bottom sector, see the second column of
Table~\ref{tab:hadronic}. The difference can be explained by the much larger
phase space in the charm sector.

The last column of Table~\ref{tab:hadronic} lists the full result, showing that 
interference effects play
an important role.  The differences between the bottom and charm sectors are 
even larger for the full result since interference between the two mechanisms is 
vastly different. 
The uncertainties for the full results are obtained by adding the uncertainties 
for the individual contributions linearly. This is done to incorporate the fact that the 
residues, $g_{DK}$ and $g_{D_s\eta}$ in the case of $\dszero$, are not necessarily 
independent quantities while they contribute largely to the uncertainties in 
the individual channels, respectively. 

Note that our results differ significantly 
from the phenomenological studies of Ref.~\cite{Faessler:2008vc}. There, the 
predicted hadronic widths for the $B^{(*)}K$ molecules are much larger, in the 
range from 50 to 90~keV. The reason is that therein  the masses of the 
molecules predicted in Refs.~\cite{Guo:2006fu,Guo:2006rp} were used, which are around 
100~MeV larger than those calculated in this paper, see Table~\ref{tab:BMasses}. We 
checked that, if we keep all the LECs to the best fit values of 
Ref.~\cite{Liu:2012zya}, and only change the subtraction constant to produce 
a mass of $B_{s0}^*$ at 5725~MeV used in Ref.~\cite{Faessler:2008vc}, then we
obtain a larger width of 73~keV, which is consistent with the result in 
Ref.~\cite{Faessler:2008vc}.

%%%%%%%%%%%%%%%%%%%%%%%%%%%%%%%%%%%%%%%%%%%%%%%%%%%%
%
\subsection{Radiative Decays}\label{sec:radiative}
\subsubsection{Power Counting}
%
%%%%%%%%%%%%%%%%%%%%%%%%%%%%%%%%%%%%%%%%%%%%%%%%%%%%
%
%%%%%%%%%%%%%%%%%%%%%%%%%%%%%%%%%%%%%%%%%%%%%%%%%%%%
%
% \subsubsection{Amplitudes}\label{sec:amplitudes}
%
%%%%%%%%%%%%%%%%%%%%%%%%%%%%%%%%%%%%%%%%%%%%%%%%%%%%
\begin{figure*}%[htbp]
%\captionsetup{width=.88\textwidth}
\centering
\includegraphics[width=\linewidth]{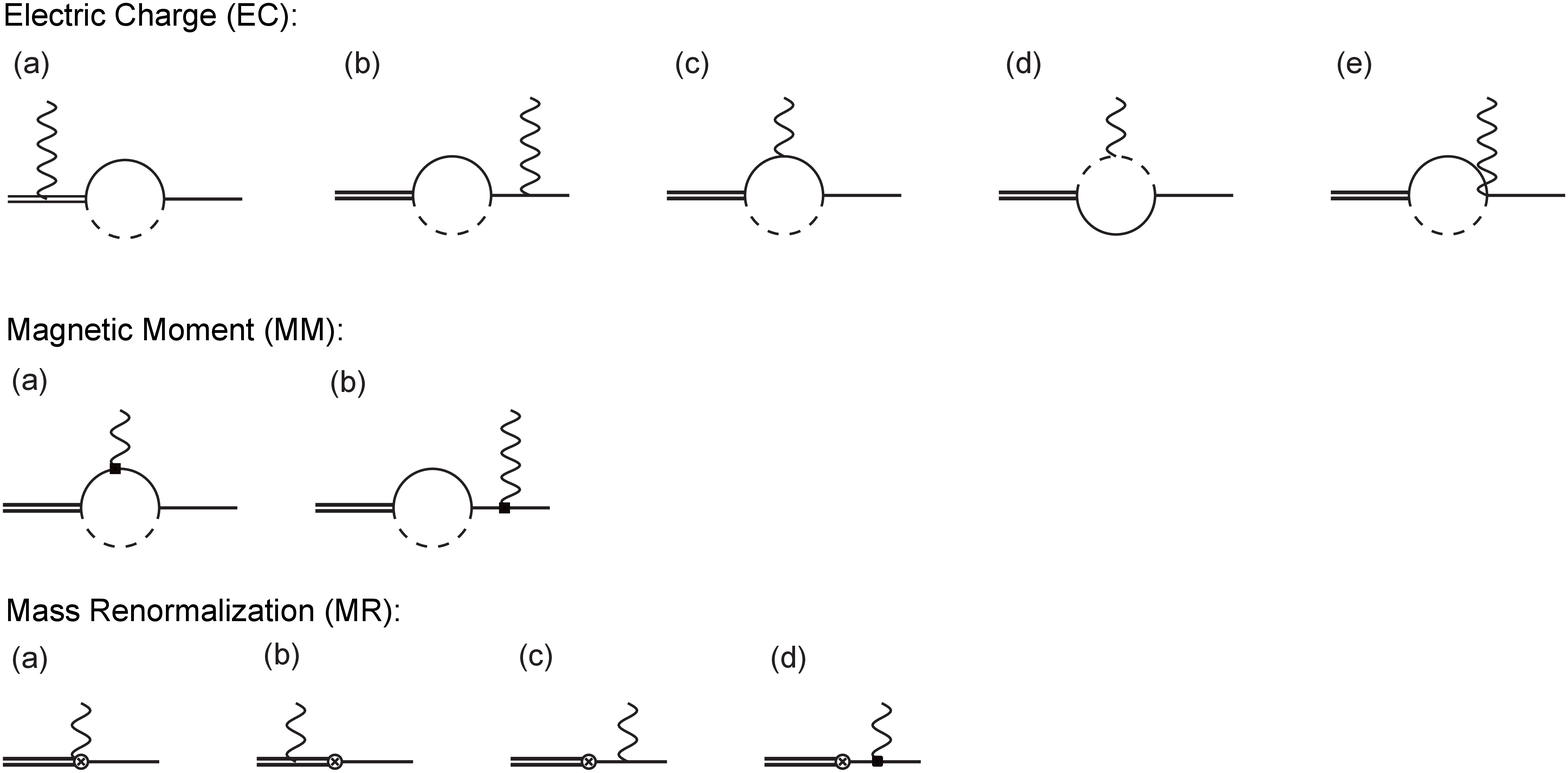}
\caption{\label{fig:diagrams} Different contributions to the radiative decays.  
In this case, the dashed lines denote kaons only. A photon coupling
to the magnetic moment is denoted by a filled box.
}\vspace{6mm}
\end{figure*}
%%%%%%%%%%%%%%%%%%%%%%%%%%%%%%%%%%%
%
%     Power Counting
%
%%%%%%%%%%%%%%%%%%%%%%%%%%%%%%%%%%%

We first address the relative size of the different contributions.  We employ
here the standard power counting scheme of CHPT coupled to heavy 
fields~\cite{Gasser:1983yg,Gasser:1984gg,Gasser:1987rb}.
The relevant momentum scale is $p\sim\sqrt{2\mu
  \epsilon}\sim m_K$. The integration measure counts as $\order{p^4}$, the 
light meson propagator as $\order{p^{-2}}$ and the heavy one as 
$\order{p^{-1}}$. Similarly, the coupling of a photon to
the electric charge gives $\order{p^2}$ for light and $\order{p}$ for heavy 
mesons. The field
strength tensor of the photon, relevant for the coupling to the magnetic
moments and the contact interaction, enters as $\order{p^2}$.  The hard scale 
in CHPT
is given by $\Lambda_\chi\sim 1\gev$. Thus, higher orders are suppressed by
positive powers in 
$p/\Lambda_\chi$.

Consider first the one-loop diagrams where the photon couples to the electric
charge of the involved mesons. For a photon emission inside the loop from a
light meson, Fig.~\ref{fig:diagrams} EC(d), we find  a factor of 
$\order{p^{-5}}$ from two light and one heavy propagators, plus the axial vector 
coupling as $\order{p}$, and the coupling of a photon to the 
electric charge of a light meson as $\order{p^2}$, so that this diagram counts 
in total as
\begin{equation}
 \order{p^4 \frac{1}{p^5}p^3} = \order{p^2}.
\end{equation}
Here, we have counted the $S$-wave coupling of the generated state to
$D^{(*)}K$ as $\order{p^0}$.  The same process with a charged intermediate heavy
meson, Fig.~\ref{fig:diagrams} EC(c), gives
\begin{equation}
 \order{p^4\frac{1}{p^4}p^2} = \order{p^2}.
\end{equation}
All other diagrams in the same set are of the same order, as required by
gauge invariance.  Since the contact interactions are proportional to the
photon field strength tensor, they also contribute at order $p^2$.  This means
that there is no enhancement of the loop diagrams compared with the contact term
and, contrary to the hadronic decays, we expect different models for the
$D_s^*$ structure to lead to similar results.  Below we will use the available
data to fix the contact interaction in one channel in order to predict the
others.

We also consider the contributions from the magnetic couplings of the heavy
mesons.  The size of the diagrams in Fig.~\ref{fig:diagrams}~MM(a) are
estimated as
\begin{equation}
\order{ p^4\frac{1}{p^4}p^3}\sim \order{p^3}.
\end{equation}
They are thus suppressed by one order in $p/\Lambda_\chi$ relative to the
LO. We will still calculate them for two reasons. First, the
explicit calculation of a higher-order correction provides a measure of how
well the chiral expansion converges. Secondly, all electric couplings vanish
for the $\dsone\to\dszero\gamma$ and $\bsone\to\bszero\gamma$ transitions, so 
that the decays via the hadron loops contribute formally only at subleading 
order, while the leading contact term $\tilde\kappa$ in Eq.~(\ref{eq:contact1}) 
enters for this channel at $\order{p^2}$. However, the corresponding loop 
is enhanced since it scales with the product of 
the residues for the
coupling of the molecules to the continuum. We therefore conclude that no
quantitative conclusions can be drawn for the $\dsone\to\dszero\gamma$ and
$\bsone\to\bszero\gamma$ decays.

Typical further higher order diagrams include an additional pion exchange in the
loop. This leads to additional factors of order $(p/\Lambda_\chi)^2$, which
imply that they can be safely neglected. The largest subleading contribution
stems from the NLO term for the axial vector coupling. It is suppressed by one
order, $p/\Lambda_\chi$, and we will use it to estimate the theoretical
uncertainty for the amplitudes.

%%%%%%%%%%%%%%%%%%%%%%%%%%%%%%%%%%%
%
%     Electric Charge
%
%%%%%%%%%%%%%%%%%%%%%%%%%%%%%%%%%%%

We close with some technical remarks.  The diagrams in the first line of
Fig.~\ref{fig:diagrams} show the full gauge invariant set of diagrams for
which the photon couples to the electric charge.  These are obtained by
gauging the kinetic terms and the axial vector coupling.  In
Ref.~\cite{thesis} the explicit expressions for the amplitudes of all possible 
transitions are given. An explicit calculation confirms that the two subsets 
($D^0K^+$ and $D^+K^0$ for $D_{s0}^*$) are gauge invariant separately, but 
there is still a remaining divergence. However, once the mass renormalization 
diagrams shown in the last line of Fig.~\ref{fig:diagrams} are included, all 
divergences are cancelled leaving a renormalized divergence-free amplitude. 

In Ref.~\cite{Lutz:2007sk} vector and axial vector states are treated in the
tensor formulation~\cite{Ecker:1989yg,Borasoy:1995ds}. In our approach, we
modify the standard treatment of the \emph{vector} formulation by employing a
trick introduced by St\"uckelberg (see, for instance, Ref.~\cite{Itzykson}). 
The standard Lagrangian for an arbitrary vector particle $V$ with mass $m_V$
reads
\begin{equation}
 \lag_V=-\frac12 V^{\mu\nu} V_{\mu\nu}^\dagger+m_V^2 V^\mu V_\mu^\dagger
\end{equation}
with the field strength tensor $V_{\mu\nu}=D_\mu V_\nu-D_\nu V_\mu$ and the
covariant derivative {$D_\mu = \partial_\mu - i\,e\,A_\mu + \ldots$}, where the
ellipses indicate the presence of additional terms not relevant 
for the following discussion and $A_\mu$ denotes the photon field. 
The propagator resulting from this Lagrangian is
\begin{equation}
 \frac{i}{l^2-m_V^2+i\varepsilon}\left(\frac{l^\mu 
l^\nu}{m_V^2}-g^{\mu\nu}\right),
\end{equation}
with $l$ the vector meson momentum.
This may make calculations rather extensive. Therefore, one 
may add another term to the Lagrangian that vanishes for 
on-shell vector particles:
\begin{equation}
 \lag_V=-\frac12 V^{\mu\nu} V_{\mu\nu}^\dagger+m_V^2 V^\mu V_\mu^\dagger-\lambda (D_\mu V^\mu)(D^\nu V_\mu^\dagger)
\end{equation}
and thus the propagator changes to
\begin{equation}
 \frac{i}{[l^2-m_V^2+i\varepsilon]}\left(\frac{1-\lambda}{\lambda}\frac{l^\mu l^\nu}{[l^2-m_V^2/\lambda+i\varepsilon]}+g^{\mu\nu}\right).
\end{equation}
where we are free to choose $\lambda=1$. Since we are dealing with radiative
decays it is important to notice that also after gauging the coupling to a
photon changes. For a much more detailed discussion of vector meson
Lagrangians and the St\"uckelberg construction, we refer the reader to the
comprehensive review~\cite{Meissner:1987ge}.

%%%%%%%%%%%%%%%%%%%%%%%%%%%%%%%%%%%
%
%     Magnetic Moment
%
%%%%%%%%%%%%%%%%%%%%%%%%%%%%%%%%%%%

\subsubsection{Results}\label{sec:results}
%
%%%%%%%%%%%%%%%%%%%%%%%%%%%%%%%%%%%
%
%     Ratios
%
%%%%%%%%%%%%%%%%%%%%%%%%%%%%%%%%%%%
\begin{table}[t]
\centering
% \captionsetup{width=0.9\textwidth}
\caption{The decay widths (in keV) calculated only from the coupling to the electric
  charge~(EC), from the magnetic moments~(MM) and from the contact term~(CT),
  respectively, compared to the total (including interference). The CT
  strength for the transitions to odd parity mesons is fixed 
  to data, while that  to even parity states, marked as '?', is undetermined
  and part of  the uncertainty.
  \medskip}
\renewcommand{\arraystretch}{1.3}
\begin{tabular}{|l|c|c|c|c|}
\hline\hline
                   Decay Channel &   EC   	&    MM		&    CT		&   Sum		\\ \hline\hline
$\dszero\rightarrow D_s^*\gamma$ & $2.0 $	& $0.03$ 	& $3.3$ 	& $9.4$		\\ \hline
$\dsone\to D_s\gamma$            & $4.2$ 	& $0.2$ 	& $11.3$ 	& $24.2$	\\ \hline
$\dsone\to D^*_s\gamma$          & $9.4$ 	& $0.5$ 	& $10.3$ 	& $25.2$	\\ \hline
$\dsone\to \dszero \gamma$       & --		& $1.3$ 	&     ? 	& $1.3$		\\ \hline\hline
$\bszero\rightarrow B_s^*\gamma$ & $22.4 $	& $0.6$ 	& $5.2$ 	& $32.6$	\\ \hline
$\bsone\to B_s\gamma$            & $39.4$ 	& $25.8$ 	& $5.1$ 	& $4.1$		\\ \hline
$\bsone\to B^*_s\gamma$          & $46.5$ 	& $0.1$ 	& $6.4$ 	& $46.9$	\\ \hline
$\bsone\to \bszero \gamma$       & --		& $0.02$ 	&     ? 	& $0.02$	\\ \hline\hline
\end{tabular}
\label{tab:results_2}
\end{table}

Our amplitudes consist of three different contributions. For the decays via
heavy meson-kaon loops, we considered the coupling of the photon to the 
electric charges and to the magnetic moment of the heavy mesons, as well as the 
contact interaction. In Table~\ref{tab:results_2}, we show the central values 
for the width calculated using one of the three contributions exclusively while 
discarding the remaining ones. With the contact interaction fixed to data as 
described below, the largest contribution comes from the loop diagrams where the 
photon couples to the electric charges.  Coupling the photon to magnetic 
moments, which is of one order higher, gives in general small contributions.  
Therefore, the chiral expansion converges well.  The magnetic contribution to 
the transition $\dsone\to\dszero\gamma$ is larger than the others in the charm 
sector, since it scales, as expected, with the product of two sizable resonance 
couplings $g_{DK}g_{D^*K}$, while all others scale with the product of one of 
them and the axial coupling constant $g_\pi$ for the $D^*D_sK$ vertex.

Two results stand out here and need to be explained. In the decay $\bsone\to
B_s\gamma$ the contribution from the magnetic moment is particularly large
because of a constructive interference that does not appear in any of the
other channels. 
Individually, the interfering diagrams do not contribute more than
expected from the power counting. However, this large contribution interferes
destructively with the electric charge contribution. This leads, in turn, to a
width about one order of magnitude smaller than those for the $\bsone\to 
B_s^*\gamma$ and $\bszero\to B_s^*\gamma$, despite of having the largest phase space.  
The same mechanism is not present in the other channels for different reasons. 
In the case of of the corresponding open charm decay, $\dsone\to D_s\gamma$, 
the charge of the heavy quark, $2/3$ instead of $-1/3$, prevents a similar effect.
For the other open bottom channels $\bszero\to B_s^*\gamma$ and 
$\bsone\to B_s^*\gamma$ the relevant loops give too small contributions.
The second interesting result is the small decay width for
$\bsone\to\bszero\gamma$.  Here we notice that the decay width scales with
$E_\gamma^3/M_{\bsone}^2$. When we compare this to the same factor in the
charm equivalent, we find a suppression of $\sim 1/150$, explaining the tiny
decay width.

\begin{table}[tb]%[htbp]
\centering
\caption{Results for the relevant decay channels, compared to PDG
  2012~\cite{Beringer:1900zz}. The $^\ast$ denotes an input quantity.
\medskip}
\renewcommand{\arraystretch}{1.3}
\begin{tabular}{|c|c|c||c|c|c|}
\hline\hline
      & Our Result	& Exp           	&       & Our Result		&     Exp          \\ \hline\hline
$R_1$ & $0.10\pm0.04$ 	&  $<0.059$		& $R_5$ & $0.98\pm0.01$ 	&  $0.93 \pm 0.09$ \\ \hline
$R_2$ & $0.38\pm0.17^\ast$ 	&  $0.38\pm0.05$	& $R_6$ & $0.37\pm0.17$		&  $0.35 \pm 0.04$ \\ \hline
$R_3$ & $0.40\pm0.16$	&  $<0.16$		& $R_7$ & $0.39\pm0.16$ 	&  $<0.24$         \\ \hline
$R_4$ & $0.02\pm0.02$	&  $<0.22$		& $R_8$ & $0.02\pm0.02$ 	&  $<0.25$         \\ \hline\hline
\end{tabular}
\label{tab:ratios}
\end{table}
The currently available data are rather limited. Only upper limits exist for
some decay widths, while others are not yet measured at all. The only
available ratios are:
\begin{eqnarray} 
 &&R_1:=\frac{\Gamma(\dszero \rightarrow D_s^*\gamma)}{\Gamma(\dszero\rightarrow 
D_s\pi^0)}, \qquad
   R_2:=\frac{\Gamma(\dsone\rightarrow D_s\gamma)}{\Gamma(\dsone\rightarrow 
D_s^*\pi^0)},\non\\ \non
 &&R_3:=\frac{\Gamma(\dsone\rightarrow D^*_s\gamma)}{\Gamma(\dsone\rightarrow 
D_s^*\pi^0)},\qquad
   R_4:=\frac{\Gamma(\dsone\rightarrow \dszero\gamma)}{\Gamma(\dsone\rightarrow 
D_s^*\pi^0)},\\ \non
 &&R_5:=\frac{\Gamma(\dsone\rightarrow D_s^*\pi^0)}{\Gamma(\dsone\rightarrow 
D_s^*\pi^0)+\Gamma(\dsone\rightarrow \dszero\gamma)},\\ \non
 &&R_6:=\frac{\Gamma(\dsone\rightarrow D_s\gamma)}{\Gamma(\dsone\rightarrow 
D_s^*\pi^0)+\Gamma(\dsone\rightarrow \dszero\gamma)},\\ \non
 &&R_7:=\frac{\Gamma(\dsone\rightarrow D^*_s\gamma)}{\Gamma(\dsone\rightarrow 
D_s^*\pi^0)+\Gamma(\dsone\rightarrow \dszero\gamma)},\\ 
 &&R_8:=\frac{\Gamma(\dsone\rightarrow \dszero\gamma)}{\Gamma(\dsone\rightarrow D_s^*\pi^0)+\Gamma(\dsone\rightarrow \dszero\gamma)}.
\end{eqnarray}

In Table~\ref{tab:ratios}, we compare our results to the experimental values.
We have chosen the ratio $R_2$ to fix the free parameter, namely the strength
of the contact interaction $\kappa$ in Eq.~(\ref{eq:contact1}). The same term 
contributes to the decays
$\dszero\to D_s^*\gamma$, $\dsone\to D_s\gamma$ and $\dsone\to D_s^*\gamma$.
However, an independent contact term $\tilde \kappa$ in 
Eq.~(\ref{eq:contact1}) contributes to $\dsone\to\dszero\gamma$. We thus assign 
an uncertainty of 100\% to this transition amplitude. The uncertainties include those 
from neglecting higher order contributions and of the 
coupling constants.  Within the theoretical uncertainties, all results agree 
with the measured ratios or upper limits. The only possible exception is $R_3$, 
which is consistent only within two
standard deviations. In our approach, almost identical values are found for
each of the pairs of ratios $R_2$ and $R_6$, $R_3$ and $R_7$ since
$\Gamma(\dsone\rightarrow D_s^*\pi^0)\gg\Gamma(\dsone\rightarrow
\dszero\gamma)$, in line with the observed proximity of $R_2$ and $R_6$.

%%%%%%%%%%%%%%%%%%%%%%%%%%%%%%%%%%%
%
%     Partial Widths
%
%%%%%%%%%%%%%%%%%%%%%%%%%%%%%%%%%%%
\begin{table}[tb]
\centering
\caption{\label{tab:results_comparison} Results for the radiative decay widths
  in keV.
  The first column gives our result with all uncertainties from higher 
orders and coupling constants added in
  quadrature. The numbers in the second column are from a parity doubling
  model by Bardeen et al.~\cite{Bardeen:2003kt}; in the third from light-cone
  sum  rules by Colangelo et al.~\cite{Colangelo:2005hv}; and in the fourth
  from Lutz and  Soyeur~\cite{Lutz:2007sk}, who provide two values with
  reasonable estimates for  their remaining free parameter. The fifth column
  reports model  calculations by Faessler et
  al.~\cite{Faessler:2007gv,Faessler:2007us,Faessler:2008vc}.
  \medskip} 
\renewcommand{\arraystretch}{1.3}
\begin{tabular}{|l|c|c|c|c|c|}
\hline\hline
Decay Channel               & Our Results   	& \cite{Bardeen:2003kt} &  \cite{Colangelo:2005hv}	& \cite{Lutz:2007sk}	& \cite{Faessler:2007gv,Faessler:2007us,Faessler:2008vc}\\ \hline\hline
$\dszero\to D_s^*\gamma$    & $(9.4\pm3.8)$	& $1.74$          	& $4-6$				& $1.94(6.47) $		&	0.55-1.41			\\ \hline 
$\dsone\to D_s\gamma$       & $(24.2\pm10.7)$ 	& $5.08$            	& $19-29 $			& $44.50 (45.14) $ 	&	2.37-3.73			\\ \hline
$\dsone\to D^*_s\gamma$     & $(25.2\pm9.7)$ 	& $4.66$            	& 
$0.6-1.1 $	 		& $21.8 (12.47)$	&	--				\\ \hline
$\dsone\to \dszero \gamma$  & $(1.3\pm 1.3)$ 	& $2.74$            	& 
$0.5-0.8$			& $0.13 (0.59)$		&	--				\\ \hline\hline
$B_{s0}\to B_s^*\gamma$     & $(32.6\pm20.8)$ 	& $58.3$            	& --				& --			&	3.07-4.06			\\ \hline
$B_{s1}\to B_s\gamma$       & $(4.1\pm10.9)$ 	& $39.1$            	& --				& -- 			&	2.01-2.67			\\ \hline
$B_{s1}\to B_s^*\gamma$     & $(46.9\pm33.6)$ 	& $56.9$            	& --		
		& --			&	--				\\ \hline
$B_{s1}\to B_{s0}\gamma$    & $(0.02\pm0.02)$ 	& $0.0061$          	& -- 		
		& --			&	--				\\ \hline\hline
\end{tabular}
\end{table}

Table~\ref{tab:results_comparison} contains the results for the individual
radiative decay widths. The theoretical uncertainties given there contain 
various contributions, we only show the sum of all added in quadrature. The 
largest uncertainty stems from the chiral expansion, which is estimated by 
multiplying the amplitudes by $(1\pm \sqrt{2\mu\epsilon}/\Lambda_\chi )$, 
followed by the uncertainty of the contact term propagated from the error of 
the data used to fix it. 
Smaller uncertainties come from the residues and the axial vector coupling $g_\pi$, determined from the pionic decay of the $D^*$. In the case of the latter improved experimental data on the $D^*$ width would be helpful.

In absence of experimental information, we can compare our results only to the 
results of previous calculations which are given in the table as
well, where the values in the last two columns were based on the hadronic 
molecular picture of the $\dszero$ and $\dsone$. Another result in the 
molecular picture was performed by Gamermann et
al.~\cite{Gamermann:2007bm}, using an flavor-SU(4) Lagrangian, with a
width of $0.475^{+0.831}_{-0.290}$~keV for the $\dszero\to D_s^*\gamma$.
  
The results by Lutz and Soyeur~\cite{Lutz:2007sk} are the closest to ours,
while other calculations generally find smaller numbers. However, the
differences are not large: all results, even those from the parity-doubling
model for the $c\bar s$ mesons~\cite{Bardeen:2003kt}, agree with ours within
two standard deviations.  Similarly, the values from different models in the
bottom sector agree within three standard deviations.   This is
however only based on our uncertainties. Once other models quote their
residual error as well, the statistical significance of any deviation will
decrease.

In contrast to this, the hadronic width is enhanced significantly for
molecular states due to an additional loop contribution, but here the leading
contact interaction is proportional to $(m_u-m_d)E_\pi$ and thus suppressed.
Consequently, calculations performed for compact $c\bar s$ states predict
significantly smaller values.   In contradistinction, the origins
of larger contributions for molecular states are the two-particle cuts in
meson loops, resulting in total widths of the order of $100\kev$, and a larger
coupling constant in Eq.~\eqref{eq:couplingsCharm}. As can be seen in
Eq.~\eqref{eq:weinberg}, the pure molecule sets an upper bound for $\lambda^2$,
with an uncertainty of $\order{R\sqrt{2\mu\epsilon}}$. In principle, the
$D^{(*)}K$ meson loops can also contribute to the width of the $c\bar s$
mesons. However, the coupling constant would be much smaller since
$\lambda^2\ll1$ for such states.
 
Similar to the charm sector, all values in the bottom sector from different 
models agree within three standard deviations, and less when the
hitherto-unknown  uncertainties of other models are considered.

%%%%%%%%%%%%%%%%%%%%%%%%%%%%%%%%%%%%%%%%%%
%
\section{Summary and Conclusion}\label{sec:summary}
%
%%%%%%%%%%%%%%%%%%%%%%%%%%%%%%%%%%%%%%%%%%
We presented hadronic and radiative decay widths of the charm-strange
resonances $\dszero(2317)$ and $\dsone(2460)$ under the assumption that they
are $D^{(*)}K$ bound states. Our results are in fair agreement with
available data. In detail, the decay widths are larger by more than one
order of magnitude for the isospin violating hadronic decays than for the
radiative decays: the hadronic widths are around 100~keV while the
radiative ones are of the order of a few keV.

Our analysis revealed that only the hadronic decays are sensitive to a
possible molecular component of both the $\dszero$ and $\dsone$--- a hadronic
width of 100 keV or larger can be regarded as a unique feature for molecular
states. This experimental accuracy could possibly be reached with 
$\overline{\text P}$ANDA at the
future accelerator facility FAIR. The origin of this enhanced width is the
presence of two-meson cuts and the large coupling constant 
of the molecules to their constitutents---those
 should be 
much smaller in for non-molecular states.  In
contrast to this, the radiative decays turn out to be similar in size for all
models for the states of interest.  From the effective field theory point of 
view, this can be understood by the presence of a counterterm at LO, which is 
of short-distance nature, in these decays.

We also made predictions for their heavy flavor partners in the open bottom
sector.  Larger binding energies lead to much reduced hadronic widths. Since
the radiative decay widths are much larger than those of the strong decays,
radiative decays appear to be the most promising discovery channels for these
positive parity bottom-strange mesons in future experiments.

%%%%%%%%%%%%%%%%%%%%%%%%%%%%%%%%%%%%%%%%%%%%%%%%%%%%%%%%%%%%%%
\section*{Acknowledgments}

HWG is particularly indebted to the Nuclear Theory group at FZ J\"ulich for
its hospitality during his Sabbatical stay.  This work was supported in part
by the US-Department of Energy under contract DE-FG02-95ER-40907 (HWG), by the
Deutsche Forschungsgemeinschaft and the National Natural Science Foundation of
China through funds provided to the Sino-German CRC 110 ``Symmetries and the
Emergence of Structure in QCD'', by the EPOS network of the European Community
Research Infrastructure Integrating Activity ``Study of Strongly Interacting
Matter'' (HadronPhysics3), and by the NSFC (Grant No. 11165005).

%%%%%%%%%%%%%%%%%%%%%%%%%%%%%%%%%%%%%%%%%%%%%%%%%%%%%%%%%%%%%%


\begin{thebibliography}{99}

%\cite{Godfrey:1985xj}
\bibitem{Godfrey:1985xj} 
  S.~Godfrey and N.~Isgur,
  %``Mesons in a Relativized Quark Model with Chromodynamics,''
  Phys.\ Rev.\ D {\bf 32}, 189 (1985).
  %%CITATION = PHRVA,D32,189;%%
  %1888 citations counted in INSPIRE as of 03 Apr 2014


%\cite{Aubert:2003fg}
\bibitem{Aubert:2003fg} 
  B.~Aubert {\it et al.}  [BaBar Collaboration],
  %``Observation of a narrow meson decaying to $D_s^+ \pi^0$ at a mass of 2.32-GeV/c$^2$,''
  Phys.\ Rev.\ Lett.\  {\bf 90}, 242001 (2003)
  [hep-ex/0304021].
  %%CITATION = HEP-EX/0304021;%%
  %659 citations counted in INSPIRE as of 03 Apr 2014


%\cite{Besson:2003cp}
\bibitem{Besson:2003cp} 
  D.~Besson {\it et al.}  [CLEO Collaboration],
  %``Observation of a narrow resonance of mass 2.46-GeV/c**2 decaying to D*+(s) pi0 and confirmation of the D*(sJ)(2317) state,''
  Phys.\ Rev.\ D {\bf 68}, 032002 (2003)
  [Erratum-ibid.\ D {\bf 75}, 119908 (2007)]
  [hep-ex/0305100].
  %%CITATION = HEP-EX/0305100;%%
  %455 citations counted in INSPIRE as of 03 Apr 2014

%\cite{Nowak:1992um}
\bibitem{Nowak:1992um} 
  M.~A.~Nowak, M.~Rho and I.~Zahed,
  %``Chiral effective action with heavy quark symmetry,''
  Phys.\ Rev.\ D {\bf 48}, 4370 (1993)
  [hep-ph/9209272].
  %%CITATION = HEP-PH/9209272;%%

%\cite{Bardeen:2003kt}
\bibitem{Bardeen:2003kt} 
  W.~A.~Bardeen, E.~J.~Eichten and C.~T.~Hill,
  %``Chiral multiplets of heavy - light mesons,''
  Phys.\ Rev.\ D {\bf 68}, 054024 (2003)
  [hep-ph/0305049].
  %%CITATION = HEP-PH/0305049;%%
  %339 citations counted in INSPIRE as of 03 Apr 2014

%\cite{Nowak:2003ra}
\bibitem{Nowak:2003ra} 
  M.~A.~Nowak, M.~Rho and I.~Zahed,
  %``Chiral doubling of heavy light hadrons: BABAR 2317-MeV/c**2 and CLEO 2463-MeV/c**2 discoveries,''
  Acta Phys.\ Polon.\ B {\bf 35}, 2377 (2004)
  [hep-ph/0307102].
  %%CITATION = HEP-PH/0307102;%%

\bibitem{Guo:2009id} 
  F.-K.~Guo, C.~Hanhart and U.-G.~Mei\ss{}ner,
  %``Implications of heavy quark spin symmetry on heavy meson hadronic 
  %molecules,''
  Phys.\ Rev.\ Lett.\  {\bf 102}, 242004 (2009)
  [arXiv:0904.3338 [hep-ph]].
  
%\cite{Barnes:2003dj}
\bibitem{Barnes:2003dj} 
  T.~Barnes, F.~E.~Close and H.~J.~Lipkin,
  %``Implications of a DK molecule at 2.32-GeV,''
  Phys.\ Rev.\ D {\bf 68}, 054006 (2003)
  [hep-ph/0305025].
  %%CITATION = HEP-PH/0305025;%%
  %272 citations counted in INSPIRE as of 03 Apr 2014


%\cite{vanBeveren:2003kd}
\bibitem{vanBeveren:2003kd} 
  E.~van Beveren and G.~Rupp,
  %``Observed D(s)(2317) and tentative D(2030) as the charmed cousins of the light scalar nonet,''
  Phys.\ Rev.\ Lett.\  {\bf 91}, 012003 (2003)
  [hep-ph/0305035].
  %%CITATION = HEP-PH/0305035;%%
  %232 citations counted in INSPIRE as of 03 Apr 2014

\bibitem{vanBeveren:2003jv} 
  E.~van Beveren and G.~Rupp,
  %``Continuum bound states K(L), D(1)(2420), D(sl)(2536) and their partners 
  %K(S), D(1)(2400), D*(sJ)(2463),''
  Eur.\ Phys.\ J.\ C {\bf 32}, 493 (2004)
  [hep-ph/0306051].

%\cite{Kolomeitsev:2003ac}
\bibitem{Kolomeitsev:2003ac} 
  E.~E.~Kolomeitsev and M.~F.~M.~Lutz,
  %``On Heavy light meson resonances and chiral symmetry,''
  Phys.\ Lett.\ B {\bf 582}, 39 (2004)
  [hep-ph/0307133].
  %%CITATION = HEP-PH/0307133;%%
  %134 citations counted in INSPIRE as of 03 Apr 2014


%\cite{Guo:2006fu}
\bibitem{Guo:2006fu} 
  F.-K.~Guo, P.-N.~Shen, H.-C.~Chiang, R.-G.~Ping and B.-S.~Zou,
  %``Dynamically generated 0+ heavy mesons in a heavy chiral unitary approach,''
  Phys.\ Lett.\ B {\bf 641}, 278 (2006)
  [hep-ph/0603072].
  %%CITATION = HEP-PH/0603072;%%
  %96 citations counted in INSPIRE as of 03 Apr 2014

\bibitem{Zhang:2006ix} 
  Y.-J.~Zhang, H.-C.~Chiang, P.-N.~Shen and B.-S.~Zou,
  %``Possible S-wave bound-states of two pseudoscalar mesons,''
  Phys.\ Rev.\ D {\bf 74}, 014013 (2006)
  [hep-ph/0604271].

%\cite{Guo:2006rp}
\bibitem{Guo:2006rp} 
  F.-K.~Guo, P.-N.~Shen and H.-C.~Chiang,
  %``Dynamically generated 1+ heavy mesons,''
  Phys.\ Lett.\ B {\bf 647}, 133 (2007)
  [hep-ph/0610008].
  %%CITATION = HEP-PH/0610008;%%
  %53 citations counted in INSPIRE as of 03 Apr 2014


%\cite{Weinberg:1962hj}
\bibitem{Weinberg:1962hj} 
  S.~Weinberg,
  %``Elementary particle theory of composite particles,''
  Phys.\ Rev.\  {\bf 130}, 776 (1963).
  %%CITATION = PHRVA,130,776;%%
  %299 citations counted in INSPIRE as of 03 Apr 2014


%\cite{Mohler:2013rwa}
\bibitem{Mohler:2013rwa} 
  D.~Mohler, C.~B.~Lang, L.~Leskovec, S.~Prelovsek and R.~M.~Woloshyn,
  %``$D_{s0}^*(2317)$ Meson and $D$-Meson-Kaon Scattering from Lattice QCD,''
  Phys.\ Rev.\ Lett.\  {\bf 111}, 222001 (2013)
  [arXiv:1308.3175 [hep-lat]].
  %%CITATION = ARXIV:1308.3175;%%
  %11 citations counted in INSPIRE as of 03 Apr 2014


%\cite{Lang:2014yfa}
\bibitem{Lang:2014yfa} 
  C.~B.~Lang, L.~Leskovec, D.~Mohler, S.~Prelovsek and R.~M.~Woloshyn,
  %``$D_{s}$ mesons with $DK$ and $D^{*}K$ scattering near threshold,''
  arXiv:1403.8103 [hep-lat].
  %%CITATION = ARXIV:1403.8103;%%




%\cite{Liu:2012zya}
\bibitem{Liu:2012zya} 
  L.~Liu, K.~Orginos, F.-K.~Guo, C.~Hanhart and U.-G.~Mei{\ss}ner,
  %``Interactions of Charmed Mesons with Light Pseudoscalar Mesons from Lattice QCD and Implications on the Nature of the $D_{s0}^*(2317)$,''
  Phys.\ Rev.\ D {\bf 87}, 014508 (2013)
  [arXiv:1208.4535 [hep-lat]].
  %%CITATION = ARXIV:1208.4535;%%
  %13 citations counted in INSPIRE as of 03 Apr 2014

\bibitem{Guo:2013nja} 
  F.-K.~Guo, L.~Liu, U.-G.~Mei\ss{}ner and P.~Wang,
  %``Tetraquarks, hadronic molecules, meson-meson scattering and disconnected 
  %contributions in lattice QCD,''
  Phys.\ Rev.\ D {\bf 88}, 074506 (2013)
  [arXiv:1308.2545 [hep-lat]].

%\cite{Guo:2009ct}
\bibitem{Guo:2009ct} 
  F.-K.~Guo, C.~Hanhart and U.-G.~Mei{\ss}ner,
  %``Interactions between heavy mesons and Goldstone bosons from chiral 
  %dynamics,''
  Eur.\ Phys.\ J.\ A {\bf 40}, 171 (2009)
  [arXiv:0901.1597 [hep-ph]].
  %%CITATION = ARXIV:0901.1597;%%
  %39 citations counted in INSPIRE as of 03 Apr 2014
  
  
%\cite{Hofmann:2003je}
\bibitem{Hofmann:2003je} 
  J.~Hofmann and M.~F.~M.~Lutz,
  %``Open charm meson resonances with negative strangeness,''
  Nucl.\ Phys.\ A {\bf 733}, 142 (2004)
  [hep-ph/0308263].
  %%CITATION = HEP-PH/0308263;%%
  %88 citations counted in INSPIRE as of 03 Apr 2014


%\cite{Faessler:2007gv}
\bibitem{Faessler:2007gv} 
  A.~Faessler, T.~Gutsche, V.~E.~Lyubovitskij and Y.~-L.~Ma,
  %``Strong and radiative decays of the D(s0)*(2317) meson in the DK-molecule picture,''
  Phys.\ Rev.\ D {\bf 76}, 014005 (2007)
  [arXiv:0705.0254 [hep-ph]].
  %%CITATION = ARXIV:0705.0254;%%
  %78 citations counted in INSPIRE as of 03 Apr 2014


%\cite{Guo:2008gp}
\bibitem{Guo:2008gp} 
  F.-K.~Guo, C.~Hanhart, S.~Krewald and U.-G.~Mei{\ss}ner,
  %``Subleading contributions to the width of the D*(s0)(2317),''
  Phys.\ Lett.\ B {\bf 666}, 251 (2008)
  [arXiv:0806.3374 [hep-ph]].
  %%CITATION = ARXIV:0806.3374;%%
  %33 citations counted in INSPIRE as of 03 Apr 2014


%\cite{Gamermann:2007bm}
\bibitem{Gamermann:2007bm} 
  D.~Gamermann, L.~R.~Dai and E.~Oset,
  %``Radiative decay of the dynamically generated open and hidden charm scalar meson resonances D(s0)*(2317) and X(3700),''
  Phys.\ Rev.\ C {\bf 76}, 055205 (2007)
  [arXiv:0709.2339 [hep-ph]].
  %%CITATION = ARXIV:0709.2339;%%
  %16 citations counted in INSPIRE as of 03 Apr 2014


%\cite{Faessler:2007us}
\bibitem{Faessler:2007us} 
  A.~Faessler, T.~Gutsche, V.~E.~Lyubovitskij and Y.~-L.~Ma,
  %``D* K molecular structure of the D(s1)(2460) meson,''
  Phys.\ Rev.\ D {\bf 76}, 114008 (2007)
  [arXiv:0709.3946 [hep-ph]].
  %%CITATION = ARXIV:0709.3946;%%
  %50 citations counted in INSPIRE as of 03 Apr 2014


%\cite{Lutz:2007sk}
\bibitem{Lutz:2007sk} 
  M.~F.~M.~Lutz and M.~Soyeur,
  %``Radiative and isospin-violating decays of D(s)-mesons in the hadrogenesis conjecture,''
  Nucl.\ Phys.\ A {\bf 813}, 14 (2008)
  [arXiv:0710.1545 [hep-ph]].
  %%CITATION = ARXIV:0710.1545;%%
  %43 citations counted in INSPIRE as of 03 Apr 2014

%\cite{Mehen:2005hc}
\bibitem{Mehen:2005hc} 
  T.~Mehen and R.~P.~Springer,
  %``Even- and odd-parity charmed meson masses in heavy hadron chiral 
  %perturbation theory,''
  Phys.\ Rev.\ D {\bf 72}, 034006 (2005)
  [hep-ph/0503134].
  %%CITATION = HEP-PH/0503134;%%
  %22 citations counted in INSPIRE as of 03 Apr 2014

\bibitem{Cheng:2014bca} 
  H.-Y.~Cheng and F.-S.~Yu,
  %``Near Mass Degeneracy in the Scalar Meson Sector: Implications for 
  %$B^*_{(s)0}$ and $B'_{(s)1}$ Mesons,''
  arXiv:1404.3771 [hep-ph].


%\cite{Cleven:2010aw}
\bibitem{Cleven:2010aw} 
  M.~Cleven, F.-K.~Guo, C.~Hanhart and U.-G.~Mei{\ss}ner,
  %``Light meson mass dependence of the positive parity heavy-strange mesons,''
  Eur.\ Phys.\ J.\ A {\bf 47}, 19 (2011)
  [arXiv:1009.3804 [hep-ph]].
  %%CITATION = ARXIV:1009.3804;%%
  %15 citations counted in INSPIRE as of 03 Apr 2014




%\cite{Gamermann:2006nm}
\bibitem{Gamermann:2006nm} 
  D.~Gamermann, E.~Oset, D.~Strottman and M.~J.~Vicente Vacas,
  %``Dynamically generated open and hidden charm meson systems,''
  Phys.\ Rev.\ D {\bf 76}, 074016 (2007)
  [hep-ph/0612179].
  %%CITATION = HEP-PH/0612179;%%
  %105 citations counted in INSPIRE as of 03 Apr 2014


%\cite{Gamermann:2007fi}
\bibitem{Gamermann:2007fi} 
  D.~Gamermann and E.~Oset,
  %``Axial resonances in the open and hidden charm sectors,''
  Eur.\ Phys.\ J.\ A {\bf 33}, 119 (2007)
  [arXiv:0704.2314 [hep-ph]].
  %%CITATION = ARXIV:0704.2314;%%
  %75 citations counted in INSPIRE as of 03 Apr 2014


%\cite{Mai:2012wy}
\bibitem{Mai:2012wy} 
  M.~Mai, P.~C.~Bruns and U.-G.~Mei{\ss}ner,
  %``Pion photoproduction off the proton in a gauge-invariant chiral unitary framework,''
  Phys.\ Rev.\ D {\bf 86}, 094033 (2012)
  [arXiv:1207.4923 [nucl-th]].
  %%CITATION = ARXIV:1207.4923;%%
  %10 citations counted in INSPIRE as of 03 Apr 2014


%\cite{Meissner:2005bz}
\bibitem{Meissner:2005bz} 
  U.-G.~Mei{\ss}ner, U.~Raha and A.~Rusetsky,
  %``The Pion-nucleon scattering lengths from pionic deuterium,''
  Eur.\ Phys.\ J.\ C {\bf 41}, 213 (2005)
  [Erratum-ibid.\ C {\bf 45}, 545 (2006)]
  [nucl-th/0501073].
  %%CITATION = NUCL-TH/0501073;%%
  %33 citations counted in INSPIRE as of 03 Apr 2014

\bibitem{thesis}
  M. Cleven, PhD thesis, University of Bonn, 2013.

%\cite{Amundson:1992yp}
\bibitem{Amundson:1992yp} 
  J.~F.~Amundson, C.~G.~Boyd, E.~E.~Jenkins, M.~E.~Luke, A.~V.~Manohar, J.~L.~Rosner, M.~J.~Savage and M.~B.~Wise,
  %``Radiative D* decay using heavy quark and chiral symmetry,''
  Phys.\ Lett.\ B {\bf 296}, 415 (1992)
  [hep-ph/9209241].
  %%CITATION = HEP-PH/9209241;%%
  %113 citations counted in INSPIRE as of 03 Apr 2014


%\cite{Hu:2005gf}
\bibitem{Hu:2005gf} 
  J.~Hu and T.~Mehen,
  %``Chiral Lagrangian with heavy quark-diquark symmetry,''
  Phys.\ Rev.\ D {\bf 73}, 054003 (2006)
  [hep-ph/0511321].
  %%CITATION = HEP-PH/0511321;%%
  %33 citations counted in INSPIRE as of 03 Apr 2014


%\cite{Beringer:1900zz}
\bibitem{Beringer:1900zz} 
  J.~Beringer {\it et al.}  [Particle Data Group Collaboration],
  %``Review of Particle Physics (RPP),''
  Phys.\ Rev.\ D {\bf 86}, 010001 (2012).
  %%CITATION = PHRVA,D86,010001;%%
  %3514 citations counted in INSPIRE as of 03 Apr 2014

\bibitem{Guo:2008ns} 
  F.-K.~Guo, C.~Hanhart and U.-G.~Mei\ss{}ner,
  %``Mass splittings within heavy baryon isospin multiplets in chiral 
  %perturbation theory,''
  JHEP {\bf 0809}, 136 (2008)
  [arXiv:0809.2359 [hep-ph]].

%\cite{Faessler:2008vc}
\bibitem{Faessler:2008vc} 
  A.~Faessler, T.~Gutsche, V.~E.~Lyubovitskij and Y.~-L.~Ma,
  %``Molecular structure of the B*(sl)(5725) and B(s1)(5778) bottom-strange mesons,''
  Phys.\ Rev.\ D {\bf 77}, 114013 (2008)
  [arXiv:0801.2232 [hep-ph]].
  %%CITATION = ARXIV:0801.2232;%%
  %32 citations counted in INSPIRE as of 03 Apr 2014


%%\cite{Scherer:2012xha }
%\bibitem{Scherer:2012xha} 
%  S.~Scherer and M.~R.~Schindler,
%  %``A Primer for Chiral Perturbation Theory,''
%  Lect.\ Notes Phys.\  {\bf 830}, pp.1 (2012).
%  %%CITATION = LNPHA,830,pp.1;%%

\bibitem{Gasser:1983yg} 
  J.~Gasser and H.~Leutwyler,
  %``Chiral Perturbation Theory to One Loop,''
  Annals Phys.\  {\bf 158}, 142 (1984).

\bibitem{Gasser:1984gg} 
  J.~Gasser and H.~Leutwyler,
  %``Chiral Perturbation Theory: Expansions in the Mass of the Strange Quark,''
  Nucl.\ Phys.\ B {\bf 250}, 465 (1985).

\bibitem{Gasser:1987rb} 
  J.~Gasser, M.~E.~Sainio and A.~Svarc,
  %``Nucleons with Chiral Loops,''
  Nucl.\ Phys.\ B {\bf 307}, 779 (1988).

%\cite{Ecker:1989yg}
\bibitem{Ecker:1989yg} 
  G.~Ecker, J.~Gasser, H.~Leutwyler, A.~Pich and E.~de Rafael,
  %``Chiral Lagrangians for Massive Spin 1 Fields,''
  Phys.\ Lett.\ B {\bf 223}, 425 (1989).
  %%CITATION = PHLTA,B223,425;%%
  %642 citations counted in INSPIRE as of 03 Apr 2014


%\cite{Borasoy:1995ds}
\bibitem{Borasoy:1995ds} 
  B.~Borasoy and U.-G.~Mei{\ss}ner,
  %``Chiral Lagrangians for baryons coupled to massive spin 1 fields,''
  Int.\ J.\ Mod.\ Phys.\ A {\bf 11}, 5183 (1996)
  [hep-ph/9511320].
  %%CITATION = HEP-PH/9511320;%%
  %35 citations counted in INSPIRE as of 03 Apr 2014

\bibitem{Itzykson}
  C.~Itzykson and J.-B.~Zuber,
  {\it Quantum Field Theory},
  McGraw-Hill, 1980.

%\cite{Meissner:1987ge}
\bibitem{Meissner:1987ge} 
  U.-G.~Mei{\ss}ner,
  %``Low-Energy Hadron Physics from Effective Chiral Lagrangians with Vector Mesons,''
  Phys.\ Rept.\  {\bf 161}, 213 (1988).
  %%CITATION = PRPLC,161,213;%%

%\cite{Colangelo:2005hv}
\bibitem{Colangelo:2005hv} 
  P.~Colangelo, F.~De Fazio and A.~Ozpineci,
  %``Radiative transitions of D*(sJ)(2317) and D(sJ)(2460),''
  Phys.\ Rev.\ D {\bf 72}, 074004 (2005)
  [hep-ph/0505195].
  %%CITATION = HEP-PH/0505195;%%
  %83 citations counted in INSPIRE as of 03 Apr 2014


\end{thebibliography}
\end{document}